\documentclass[preprint,times]{aastex631}
\usepackage{listings}
\usepackage{times}
\usepackage{xpatch}
\usepackage{graphicx}
\usepackage{setspace} 
\xpatchcmd{\thebibliography}{\twocolumngrid}{}{}{}

\begin{document}
\newcommand{\vdag}{(v)^\dagger}
\newcommand\aastex{AAS\TeX}
\newcommand\latex{La\TeX}
\newcommand*\mean[1]{\bar{#1}}nd\latex{La\TeX}

\title{`Weather' in the LSST Camera: Investigating Patterns in Differenced Flat Images}

\author[0000-0002-0786-7307]{John Banovetz}
\affiliation{Brookhaven National Laboratory, 20 North Technology St, Upton, NY, 11973, USA}

\author[0000-0001-6161-8988]{Yousuke Utsumi}
\affiliation{SLAC National Accelerator Laboratory, 2575 Sand Hill Road, Menlo Park, CA 94025, USA}

\author[0000-0002-2308-4230]{Joshua Meyers}
\affiliation{SLAC National Accelerator Laboratory, 2575 Sand Hill Road, Menlo Park, CA 94025, USA}

\author[0000-0003-1285-8170]{Maya Beleznay}
\affiliation{SLAC National Accelerator Laboratory, 2575 Sand Hill Road, Menlo Park, CA 94025, USA}
\affiliation{Department of Physics, Stanford University, Stanford, CA 94305}

\author[0000-0001-5738-8956]{Andrew Rasmussen}
\affiliation{SLAC National Accelerator Laboratory, 2575 Sand Hill Road, Menlo Park, CA 94025, USA}

\author[0000-0001-5326-3486]{Aaron Roodman}
\affiliation{SLAC National Accelerator Laboratory, 2575 Sand Hill Road, Menlo Park, CA 94025, USA}




\begin{abstract}
During electro-optical testing of the camera for the upcoming Vera C. Rubin Observatory Legacy Survey of Space and Time, a unique low-signal pattern was found in differenced pairs of flat images used to create photon transfer curves, with peak-to-peak variations of a factor of $10^{-3}$. A turbulent pattern of this amplitude was apparent in many differenced flat-fielded images. The pattern changes from image to image and shares similarities with atmospheric `weather' turbulence patterns. We applied several strategies to determine the source of the turbulent pattern and found that it is representative of the mixing of the air and index of refraction variations caused by the internal camera purge system displacing air, which we are sensitive to due to our flat field project setup. Characterizing this changing environment with 2-D correlation functions of the `weather' patterns provides evidence that the images reflect the changes in the camera environment due to the internal camera purge system.
Simulations of the full optical system using the \texttt{galsim} and \texttt{batoid} codes show that the weather pattern affects the dispersion of the camera point-spread function at only the one part in $10^{-4}$ level.
\end{abstract}



\section{Introduction} \label{sec:intro}


The Legacy Survey of Space and Time (LSST) to be conducted at Vera C. Rubin Observatory and set to begin in 2025 will revolutionize our understanding of dark energy and dark matter, transient events like supernovae, and even our own solar system \citep{Ivezic2019}.
The LSST camera (LSSTCam), the largest digital camera ever constructed, with a 740\,mm diameter focal plane populated with 189 4k$\times$4k science charge-couple device (CCD) sensors plus 12 guide and wavefront sensing CCDs \citep{Ivezic2019} was integrated and tested over several years at the SLAC National Accelerator Laboratory.  The fully integrated camera was characterized via electro-optical (EO) testing at SLAC in 2023 prior to the shipment of LSSTCam to Rubin Observatory on Cerro Pachon in Chile. The pre-shipment testing using flat image and spot projectors provided important diagnostics for all 201 sensors that will be invaluable to minimize systematics once the survey is underway.


Earlier EO testing runs with LSSTCam at various stages of integration have already provided detailed information to characterize the sensors and electronics \citep{Kotov2016,OConnor2016,Lopez2018,Newbry2018}. These setups allowed for studies of the CCD bright-fatter effect \citep{Broughton2024}, sensor systematics in the measurements  \citep{Snyder2020,Esteves2023}, and sensor characteristics \citep{Synder2021}. More recent results can be found in \cite{Roodman2024} and \cite{Utsumi2024}.  The 2023 series of EO tests was the only one with the camera fully integrated, in particular with the three-element lens system installed and the camera body panels along with its associated internal air handling ducting installed.


A central diagnostic from EO testing is the photon transfer curve (PTC; \cite{Janesick2001}), which is derived from pairs of flat images obtained for a range of integrated fluxes of light and is used to measure the gain and read noise of each of the sensors as well as provide information to measure the effective pixel size  and the pixel well capacity \citep{Astier2019,Broughton2024}. 
When we first constructed a PTC for the integrated LSSTCam, we calculated the pixel correlations using methods similar to \cite{Astier2019} and noticed there was unexpected degree of pixel correlation between different pixel pairs, with correlations extending roughly two to four times higher than expected. While we expected the pixel correlations to drop to a level around $10^{-3}$ after 100 microns=10 pixels (corresponding to the depth of the individual pixel). However, pixel correlations in the latest electro-optical testing run plateaued above $10^{-3}$ at ranges greater than 40 pixels.
Investigating the flat images with equal illumination across the focal plane, used to create these curves, showed structure at the scale of CCDs and smaller when differenced.
We found a turbulent weather-like pattern in difference images that is persistent at low levels ($10^{-3}$) in high-flux (half full-well) flat images  across the entire focal plane.



In this paper, we describe our discovery of the `weather' phenomenon in LSSTCam data and discuss the effect, if any, that it will have on the point-spread function (PSF) of the full optical system. In Section 2 we describe the image acquisition process and the various kinds of of flat field images that were taken to study this effect. The turbulent structure across the full focal plane is shown and described in Section 3 as well as our characterization of this structure. Section 4 discusses the possible reasons for this structure and simulate the phenomenon to determine the effect it will have on the PSF with the simulated full optical system created via \texttt{galsim} and \texttt{batoid} \citep{Rowe2015,Batoid}. We conclude in Section 5. 


\begin{figure}[!tp]
    \centering
    \includegraphics[width=0.85\textwidth]{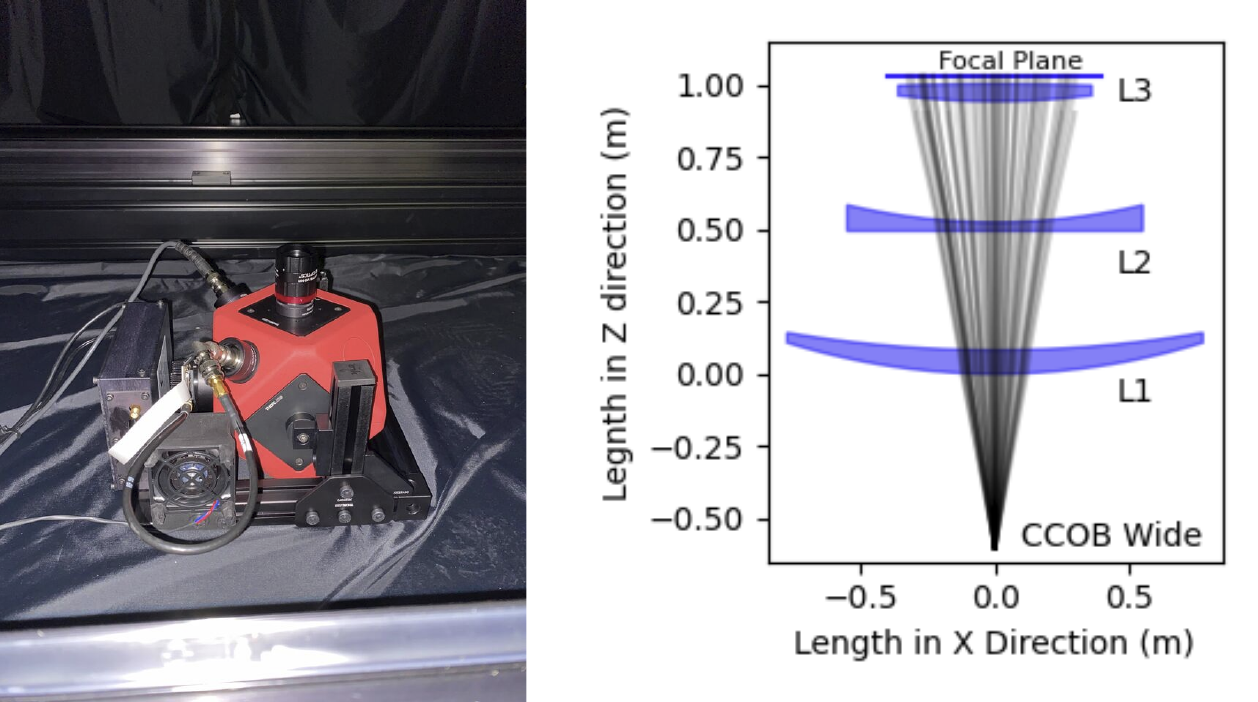}
    \caption{\textbf{Left:} The CCOB Wide Beam projector in the dark box underneath LSSTCam on the test bench. \textbf{Right:} The optical path from the projector, through the first lens (L1), then the second lens (L2), then the filter (for all images discussed in this paper, no filter was used), and the CCD window (L3), and finally to the focal plane.}
    \label{fig:CCOB_Wide}
\end{figure}

\section{Image Acquisition and Description} \label{sec:style}

\begin{figure}[!tp]
    \centering
    \includegraphics[width=0.98\textwidth]{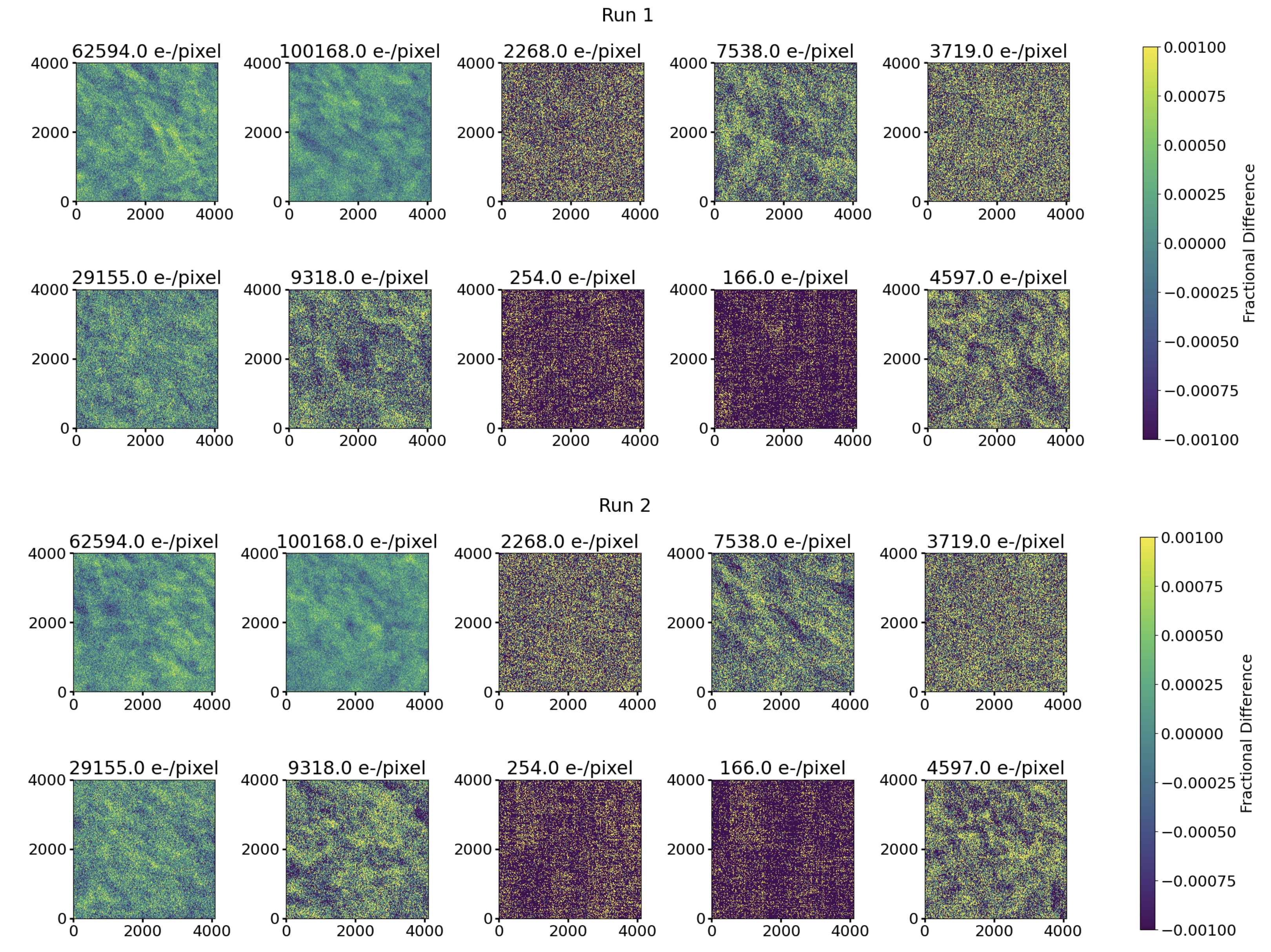}
    \caption{Two different PTC datasets that highlight the weather pattern. Each shows unbinned difference images of varying light intensity from the CCOB Wide Beam for a single CCD. The images are displayed in acquisition order from left to right. The PTCs were acquired by randomly ordering the targeted signal, shown above each of the difference images. The weather pattern changes from image to image and between datasets. The pattern is also most apparent in the higher-flux images.}
    \label{fig:PTC_Weather}
\end{figure}


The latest round of EO testing, conducted in June 2023, utilized two new projectors:  the Camera Calibration Optical Bench (CCOB) Wide Beam and Narrow Beam \citep[see][for details]{Roodman2024}. All of the images that we discuss and analyze in this paper were taken with the Wide-beam projector and were reduced with the LSST Science Pipelines \citep{LSSTPipelines,Plazas2024}.



Figure \ref{fig:CCOB_Wide} shows the CCOB Wide Beam projector in the dark box. The CCOB Wide Beam projector has an integrating sphere that uses six LEDs to approximate the bandpasses of the six LSST filters \citep{Utsumi2024}. The integrating sphere also contains a NIST-calibrated Hamamatsu photodiode that is used for calibrating the light sources. The light exits the integrating sphere through a 2.5\,cm $f/2.5$ lens to illuminate the focal plane, a slower beam than the converging $f/1.2$ beam of the Rubin observatory telescope. For EO testing, the integrating sphere was centered within a few millimeters of the center of the focal plane and 60\,cm from the first lens (L1). 
Figure \ref{fig:CCOB_Wide} also shows the light path of the photons coming from the projector and onto the focal plane.

\begin{figure}[!htp]
    \centering
    \includegraphics[width=0.98\textwidth]{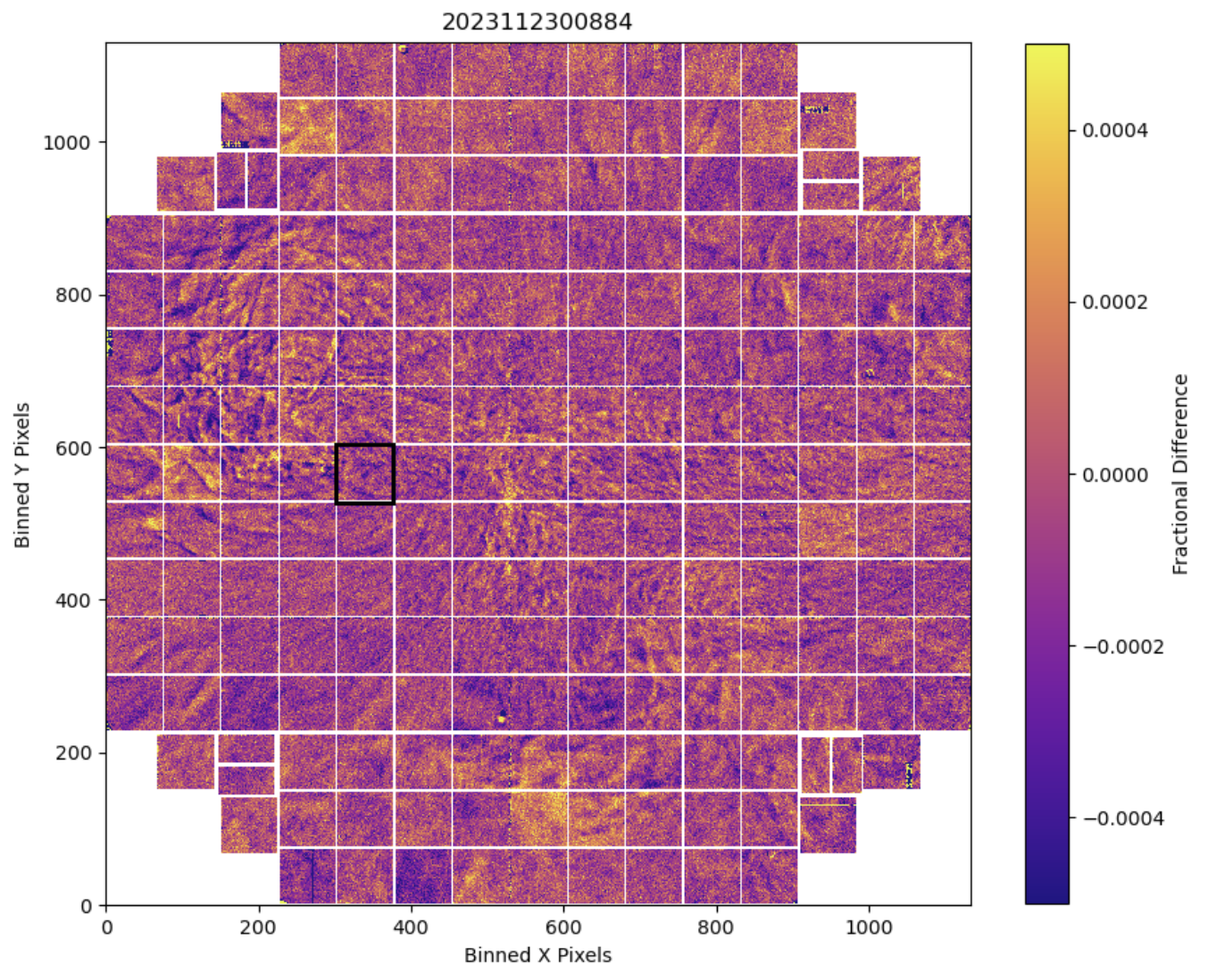}
    \caption{Example of a focal plane stability sflat image derived from Eqn.~\ref{eqn:weather} The axes represent the pixel space of the 56$\times$56 binned images, with the gaps in between the detectors still visible. The title shows the Exposure ID of the image. The color scale shows the fractional variation of the particular exposure with respect to the overall average stability flat image. The CCD used for the PTC and 2-D correlation analysis is highlighted in the black box. 
    The swirling features are the weather-like pattern for this image.}
    \label{fig:FP_Ex}
\end{figure}

The flat images that are used for making a PTC are taken in pairs ranging in brightness from $100 e^{-}$/pixel to past full well ($>1.1\times10^5 e^{-}$/pixel). The flat pairs are taken sequentially with a bias exposure (i.e. a zero exposure time image) taken in between flat pairs. A contrasting coherent structure appeared in difference normalized images, created by subtracting one of the PTC flat pair exposures at a given flux from the other. This pattern was dubbed the `weather' as it resembles atmospheric turbulence in on-sky images. Figure \ref{fig:PTC_Weather} shows PTC flat images for a range of flux levels on sensor R21/S21 (highlighted in black in Figure \ref{fig:FP_Ex}), one of the Teledyne E2V CCD detectors \citep[see][for details]{Utsumi2024}. The weather pattern is more pronounced at higher fluxes ($>50000 e^{-}$/pixel), at which it is not shot noise dominated. 


We collected a large number of flat images at a high flux level of $50,000 e^{-}$/pixel (sflats). We use this data set for assessing how the weather pattern evolves over the time.
These images were taken with the CCOB red LED (which spectrally peaks within the $r$-band filter for LSST) that flashed for roughly 0.9\,s during the 15\,s integration time. We obtained two types of sflat images, which we will denote as standard and stability sflats, distinguished by their acquisition cadence. Standard sflats were acquired sequentially, with 10--15 images with 15\,s integrations followed by approximately 3\,s readout.
The stability flats followed the same 15\,s integration time with a 3\,s readout but included a 60\,s delay in between acquisitions over a 24-hour period ($\sim$1100 images). 
We use both sflat datasets to assess how the weather pattern evolves temporally.

\section{Full Focal Plane Weather Patterns} \label{sec:focal_plane_weather}



\begin{figure}[!htp]
    \centering
    \includegraphics[width=0.98\textwidth]{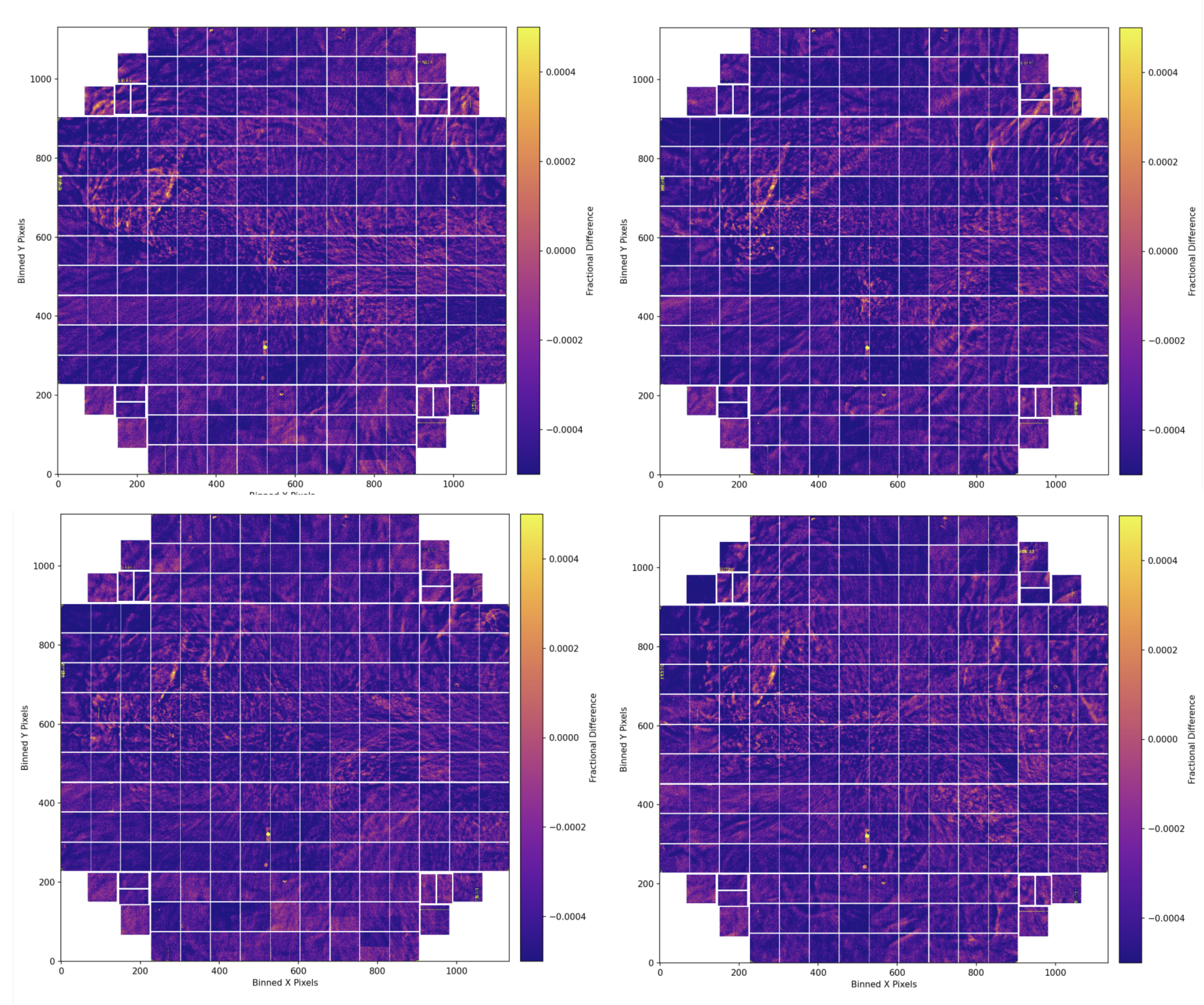}
    \caption{Similar to Figure \ref{fig:FP_Ex} but displaying consecutive images, with one minute delay between images. This highlights both the changing nature of the weather pattern as well as the presence of quasi-static non-zero fractional difference features, most likely due to dust or condensation.}
    \label{fig:FP_Multi}
\end{figure}

While the weather pattern was first noticed in PTC images, this study will focus on sflat images. We prefer these to the PTC images as the weather pattern is always easily detectable in these images and the constant flux allows us to be certain that the changes are solely due to the weather effect. Figure \ref{fig:FP_Ex} shows an example of the weather effect across the full focal plane for a stability sflat image. This image is downsampled with 56$\times$56 pixel binning in which an individual flat is subtracted by an average combined flat of all the sflats taken during the run (in this case, all the images over the 24 hour stability dataset) and then divided by the combined flat. Mathematically this is represented as
\begin{equation}
    I_{weather, i}=\frac{I_{i}-\mean{I}}{\mean{I}}
    \label{eqn:weather}
\end{equation}
where $I_{i}$ is the individual image, $\mean{I}$ is the average of the entire set of images, and $I_{weather, i}$ is the resulting `weather' image for this exposure, as seen in Figure \ref{fig:FP_Ex} and for multiple exposures in Figure \ref{fig:FP_Multi}. As the weather pattern structures appear randomly across the focal plane, 
the final average of 1100 images should be the weather-free reference flat because it only contains a few percent of the original weather component.  An animation of the entire run is available\footnote{\url{https://drive.google.com/file/d/1pJRsawh8nK0ljmXzeI1UO9sHmLImqlud/view?usp=sharing}}. The animation shows a slight drift in the global flux of the focal plane which is due to a drift in the gain of the detectors at the $10^{-4}$ level \citep[see][for details]{Roodman2024}.


\subsection{Characterizing The Weather with 2-D Correlation Functions} 
We investigated the changing pattern on shorter timescales with the sequential standard sflat images. One possible cause of the structured pattern is air turbulence due to the Volume Purge System in LSSTCam, otherwise known as the Volume Purge Cabinet (VPC). It blows clean, dry air across the lenses to ensure that no frost develops on them (the CCDs are operated at cryogenic temperatures). We experimented with different fan speeds along with turning the VPC off to investigate the effect it has on the patterns. 



\begin{figure}
    \centering
    \includegraphics[width=0.98\textwidth]{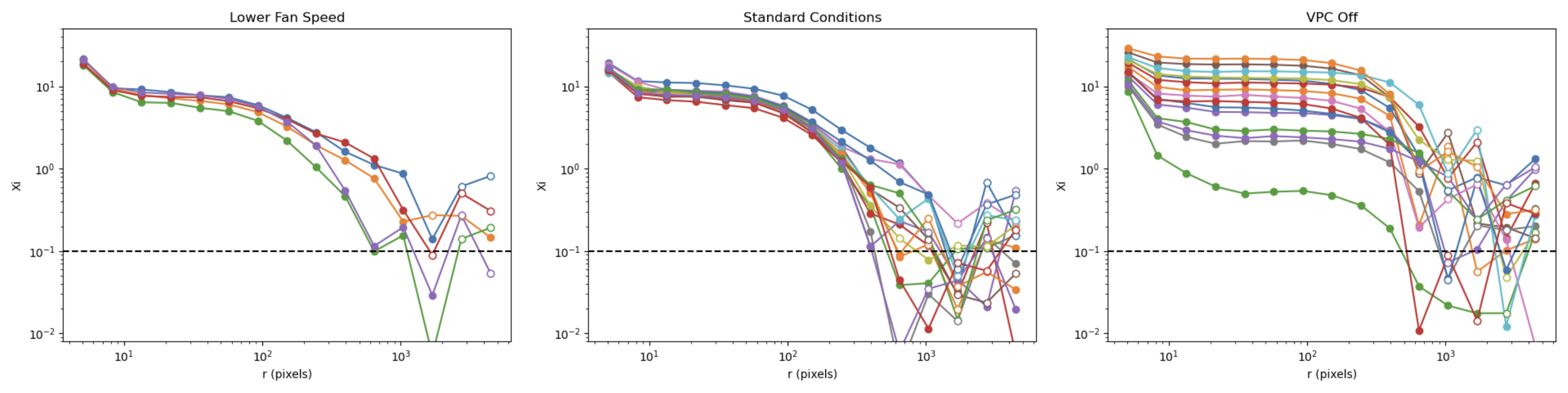}
    \caption{Correlation functions of standard sflat images of a single CCD taken with different conditions: (Left) with the VPC on and a lower fan speed (roughly 7100 rpm); (Middle) with the VPC on at standard speed (roughly 7900 rpm); (Right) with VPC off. Each of the curves represents a different exposure in the standard sflat set with positive correlations represented by a filled marker and negative with an empty marker and the black dashed line corresponds to the maximum amplitude of the 2-D correlation function of a bias image (representative of electronic noise in the detector). Only five exposures were taken at the lower fan speed. The correlation function is markedly different with the VPC off, as is the overall pattern (Figure \ref{fig:FP_No_VPC}). The increase in the length of the plateau of the 2-D correlation is likely due to the large streak-like patterns that can change across the focal plane over time (see Figure \ref{fig:FP_No_VPC}). This change in correlation function provides evidence that the weather patterns are due to turbulence in the air inside the camera body.}
    \label{fig:Xi_VPC_v_noVPC}
\end{figure}



\begin{figure}[!htp]
    \centering
    \includegraphics[width=0.98\textwidth]{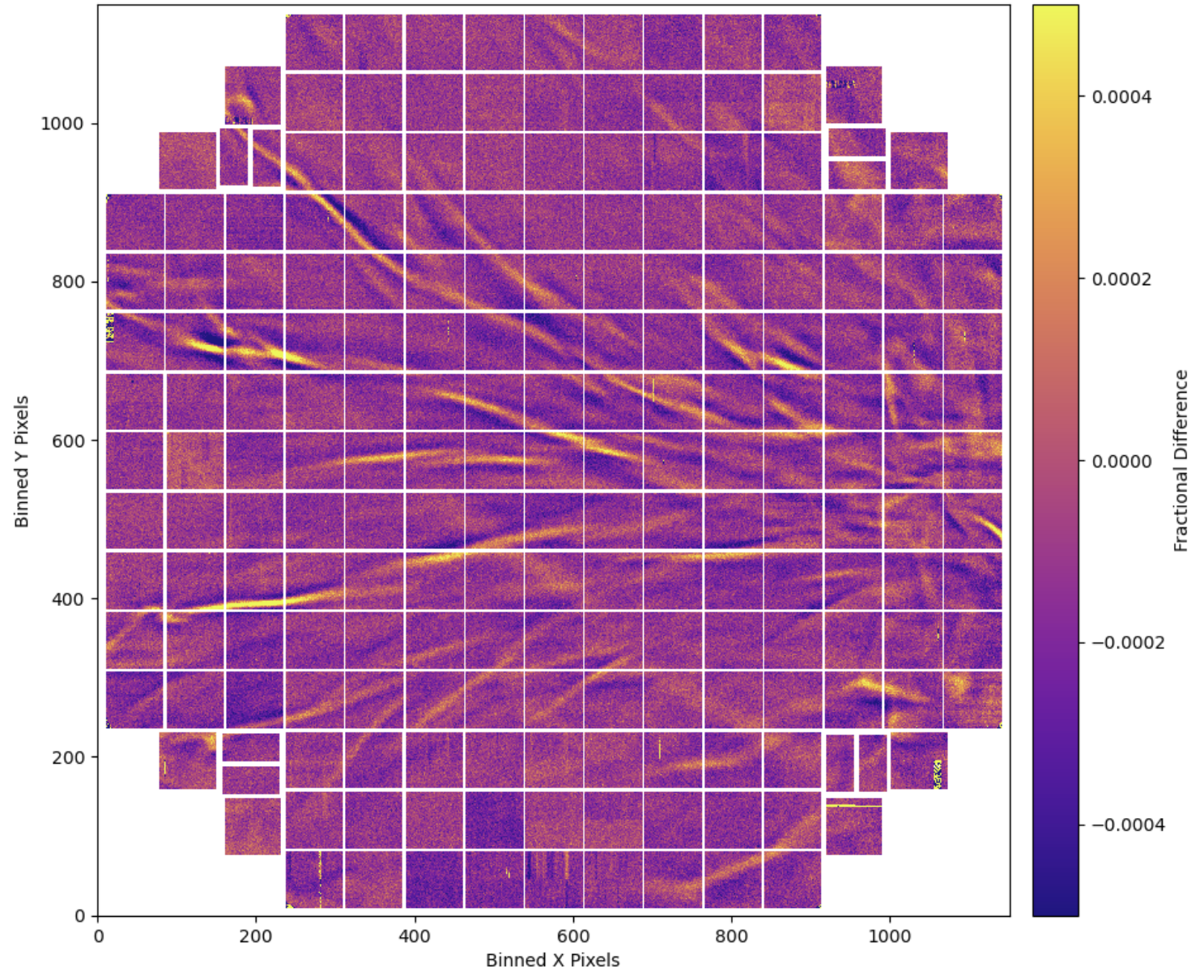}
    \caption{Similar to Figure \ref{fig:FP_Ex} but displaying an image taken with the VPC off. This shows an entirely different structure than with it on, showing stream-like structures of fractional differences across the focal plane. The difference of structure in the `weather' compared to when the VPC is turned on provides evidence that this is not a sensor level effect, especially as this new structure has `stream'-like patterns not seen with the VPC on. These streams originate to the right of this diagram, consistent with a filter exchange component being warmer than the ambient air.}
    \label{fig:FP_No_VPC}
\end{figure}

To investigate the origin of the `weather', we analyzed image data for three different settings of the VPC fan speed:  the standard speed, a lower speed, and entirely off. 
To characterize the weather patterns in these different scenarios, we investigated the 2-D correlation functions of individual images to find any differences using standard sflats. 
To calculate the 2-D correlation functions, we utilized the package \texttt{treecor}\footnote{\url{https://rmjarvis.github.io/TreeCorr/_build/html/index.html}}, specifically the \texttt{KKCorrelation} module \citep[see][for details on the algorithm]{Jarvis2004},
and ran it on $4\times4$ binned unnormalized difference images of an individual CCD located to the left of center of the focal plane shown in Figure \ref{fig:FP_Ex}, where the weather pattern appears most frequently.
The resulting 2-D correlation functions are shown in Figure \ref{fig:Xi_VPC_v_noVPC}. 
All three scenarios show a plateau feature out to $r>100$ pixels, which is not expected as correlations should decrease the as $r$ increases.
And while there are only slight variations in the correlation functions measured for images taken with the different fan speeds, the difference between turning the VPC on/off shows remarkable change in the correlations, as can be seen in the characterization in Figure \ref{fig:Xi_VPC_v_noVPC}. 
In Figure \ref{fig:FP_No_VPC} (VPC Off), the structure of the weather itself is seen to have changed to stream-like features across the focal plane. These streams vary temporally and can appear and disappear in individual detectors, explaining the wide range of values seen in Figure \ref{fig:Xi_VPC_v_noVPC}. 



Additional evidence that the `weather' effect is influenced by the VPC system comes from the VPC design. 
The VPC forces air through nozzles to generate the air flow in the camera body. These nozzles are placed on three sides of the focal plane, the top, bottom, and left side of Figures \ref{fig:FP_Ex} and \ref{fig:FP_No_VPC}. The right side does not have any nozzles as that is where a component of the filter exchange system is located. This component could be warmer than the surrounding camera body air.
Evidence for this is seen in Figure \ref{fig:FP_No_VPC} where the majority of the turbulent structure originates from the right. While some stream-like features spread throughout the focal plane, the consistently affected area is the right side of the focal plane. This suggests that the difference patterns in Figure \ref{fig:FP_Ex} are due to the turbulence of the air inside the camera body caused by the VPC, possibly close to the L3 where the VPC nozzle exists.


Another question is why LSSTCam is sensitive to the turbulent air in the camera body. Strong evidence suggests that the reason is our illumination setup. For digital astronomical cameras, flat field images are usually obtained using an illuminated screen (dome flats) or via imaging the sky near twilight (twilight flats). Both of these methods rely on converging light being incident on the focal plane. However, our method with the CCOB Wide Beam projector, the size and location of which are constrained by limitations of the test set up, uses diverging light in order to illuminate the entire focal plane. 
We also positioned the CCOB projector near the first lens, L1, greatly increasing the focal ratio of the system, much larger than it will be in operation.
The diverging optical geometry increases the sensitivity of the setup to minor air density fluctuations in the camera body (in between the light source and the detectors), to alter the path of the light, so that the effect is amplified once the light reaches the detectors. This setup is similar to that of a pinhole filter, where each pixel sees a small solid angle of a light source and any small changes (such as the index of refraction) will cause variations in the flux contrast. The weather in our setup is further enhanced by the flash time of the LEDs. With a less than one second flash time, air density variations cannot be averaged out like they would be in a much longer illumination period. Though applying this analysis to previous EO testing persiods with much longer flash times (order of 100 seconds), we still see the weather effect but at a decreased level (roughly peaking at $10^{-4}$), so the flash time is not the only contributor to this effect. The combination of diverging light, large focal ratio number, and short flash times, leads to detectable turbulence effects in the EO test flat images.

\section{`Weather' with Full Camera setup using \textit{batoid} and \textit{galsim}}

\begin{figure}[!htp]
    \centering
    \includegraphics[width=0.98\textwidth]{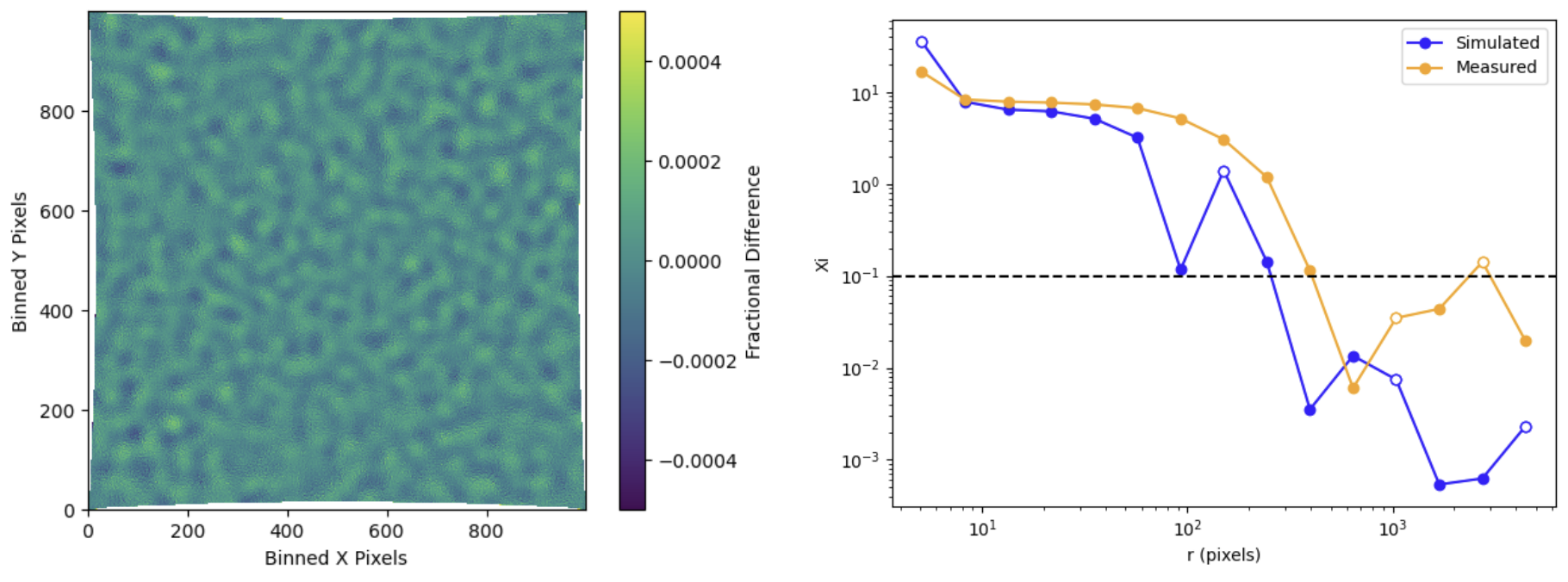}
    \caption{\textbf{Left} Example of a simulated fractional difference image derived from two images created by \texttt{batoid}. One of the images contains the weather phase screen (representing a single exposure in Equation \ref{eqn:weather} while the other doesn't (representing the combined flat image in \ref{eqn:weather}). The image is bowed as we are inputting a square phase screen onto the circular camera optics, resulting in projection effects. \textbf{Right} The 2-D correlation function of the simulated difference image compared with the correlation function of a real image taken with standard conditions, with positive correlations denoted with a solid points, negative correlations with an empty points, and the black dashed line corresponds to the maximum amplitude of the 2-D correlation function of a bias image. Both the difference image and correlation functions are used as tests to verify that the peaks and valleys and general shape of the 2-D correlation function (strength and plateau) are consistent with those of the measured difference images.}
        \label{fig:Sim_Diff_Xi}
\end{figure}



We assess to first order the impact of the `weather' on the PSF for the operation of the camera during the LSST survey, including the full optical setup and f/number of the telescope.
We utilized the packages \texttt{galsim} and \textit{batoid} to simulate light from stars going through the entire optical system and measure the PSF with and without weather effects simulated via phase screen. To roughly match the real images, we first assess if the phase screen can create images that match the peak-to-peak variations seen in the actual images, as well as the plateau seen in the 2-D correlations.

The first issue is that the `weather' pattern captured in a flat image is a part of the information of the subsection of the optical path because of the geometry of the light beam. 
The weather pattern observed in images obtained with the CCOB projector is obtained with diverging light at the focal plane. We needed to deproject it or use some alternative approach. Since we do not know the exact displacement of the density variations of air from the focal plane (the `weather' is of course an integrated effect over the line of sight), deprojection is not straightforward. Instead we make an assumption that a single phase screen can represent the weather. 
We utilize \texttt{batoid} to find the appropriate phase screen. To match the phase screen, we started with simulating the optical system of just the lenses (similar to EO testing and Figure \ref{fig:CCOB_Wide}). To mimic the weather pattern observed in CCOB images, we created a Gaussian random field of density variations utilizing a power spectrum $P(k)$ as a function of wavenumber $k$, characteristic of turbulence. 

\begin{equation}
    P(k)=Ak^{2}\left (1+\left(\frac{k}{k_{c}}\right)^{5/2} \right )^{-1}
\end{equation}

\begin{figure}[!tp]
    \centering
    \includegraphics[width=0.98\textwidth]{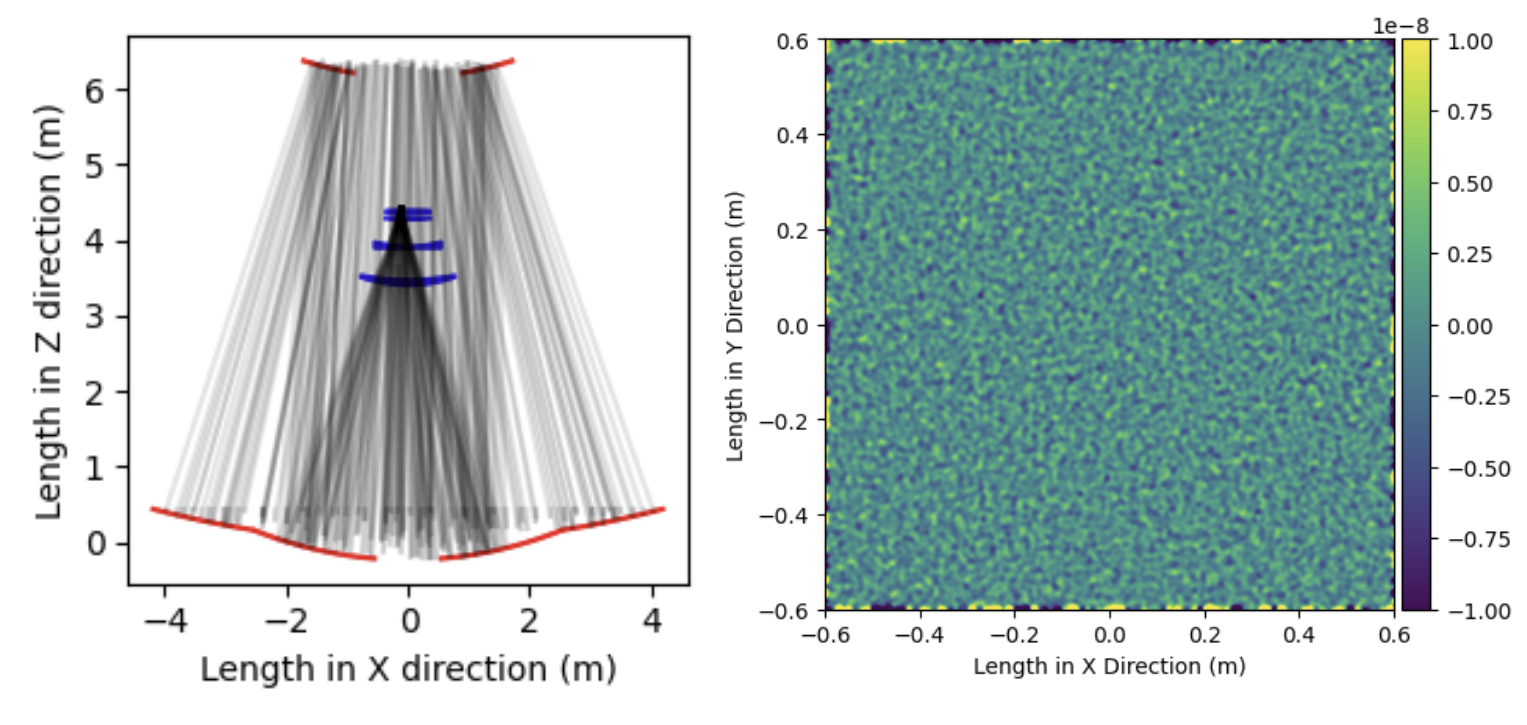}
    \caption{\textbf{Left:} Illustration of photons from a star going through the optical path of the full telescope system, with the mirrors in red, the lenses in the camera body in blue, and the weather phase screen in yellow. The star is offset from the center of the focal plane.  \textbf{Right:} An example of a Gaussian random field that we used as the weather phase screen in \texttt{batoid}. This is placed between L1 and L2 in the simulation. }
    \label{fig:Trace_Full}
\end{figure}

\begin{figure}[!tp]
    \centering
    \includegraphics[width=0.98\textwidth]{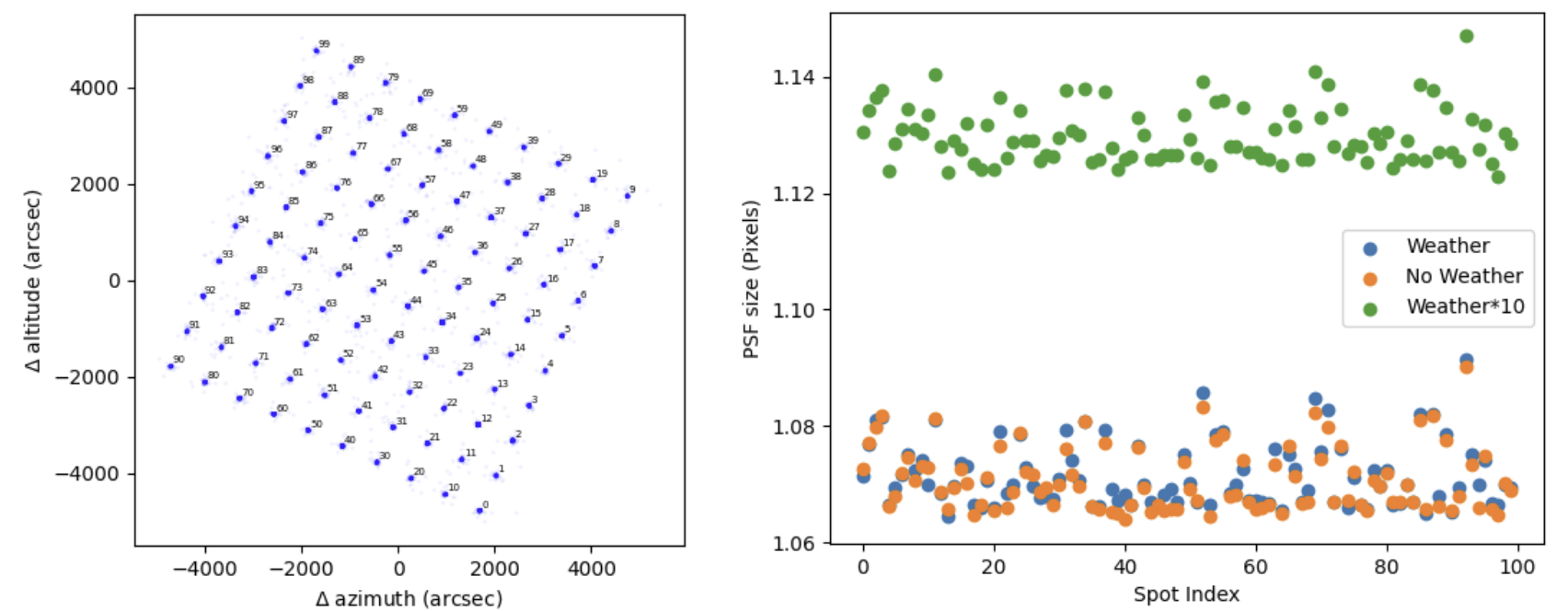}
    \caption{\textbf{Left:} The locations of the 100 simulated stars relative to the center of the focal plane.  The stars, generated with \texttt{galsim}, were input for \texttt{batoid}. \textbf{Right:} PSF dispersion (sigma) for each of the three scenarios and each of the 100 stars. There is no discernible difference across the 100 stars as well as between the normal weather phase screen and the no weather phase screen. The simulations with a stronger weather phase screen do result in a general increase of dispersion that is consistent with what we find when we measure a single simulated star 100 times.}
        \label{fig:Galsim_Field}
\end{figure}

This Gaussian random field is then scaled to and smoothed with Gaussian smoothing to create the final phase screen. 
We generated simulated CCOB flat images including the phase screen in the simulation, with the screen located in between L1 and L2. 
We then calculated the difference image between  a simulated flat image with the weather phase screen and one without the phase screen. We varied the parameters of the power spectrum until both the peaks and valleys and general shape of the 2-D correlation (primarily the plateau of $5<r<100$ pixels) of the simulated difference image were generally consistent with the  measured difference images. An example of a simulated difference image and 2-D correlation function can be seen in Figure \ref{fig:Sim_Diff_Xi}. 
We found that $A = 0.5$ pix$^{-3}$ and $k_{c} =5$ pix$^{-1]}$ produce a reasonable representation of a simulated difference image and 2-D correlation as compared to those of measured flat images. 


Once we finalized the phase screen, we added in the mirrors and an idealized atmospheric phase screens using \texttt{galsim} to simulate the full optical system (see Appendix for \texttt{galsim} parameters to create the atmosphere and the stars). Figure \ref{fig:Trace_Full} shows the ray tracing of the full system as well as an example of the final `weather' phase screen used. We then created 100 diffraction-limited stars across the focal plane and compared the PSF sizes of the 100 different stars across the focal plane with one realization of the atmosphere and one weather instance.  We also measured one simulated star 100 times with randomized instances of weather and atmosphere phase screens. We did this for three test cases: with the weather phase screen, without the phase screen, and with a weather phase screen amplitude increased by a factor of 10 to exaggerate the weather pattern. To create the stellar sources, we utilized \texttt{galsim} to simulate the light paths of point sources through the atmosphere to positions just before the telescope, and then used those photon positions as input for a \texttt{batoid} simulation. Then after computing a ray trace of the photons through the camera body optical system  from \texttt{batoid}, we use \texttt{galsim} to create a pixelated image and measure the PSF dispersion using \texttt{galsim.hsm.FindAdaptiveMom}, which is a \texttt{galsim} implementation to measure brightness moments using adaptive elliptical gaussian weights \citep{2003MNRAS.343..459H,Mandelbaum2005}.

\begin{figure*}[!tp]
    \centering
    \includegraphics[width=0.98\textwidth]{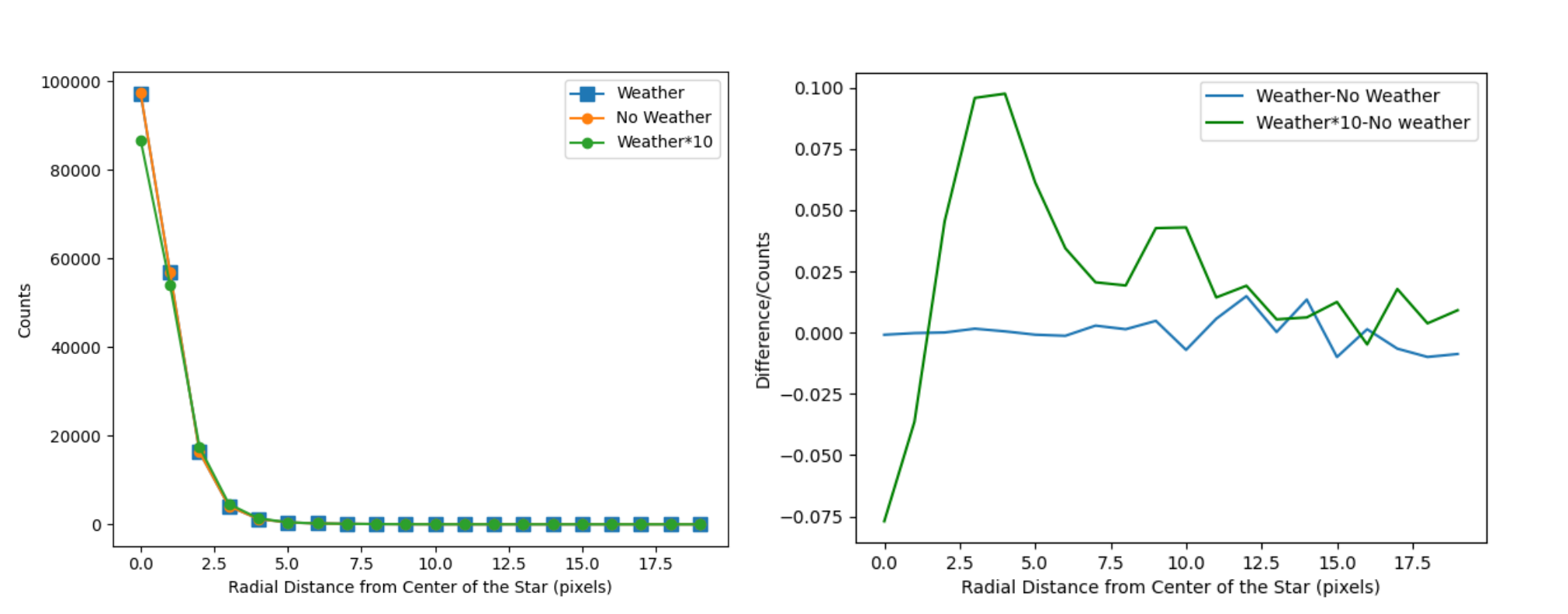}
    \caption{\textbf{Left:} Average radial profile combining all instances of a single star measured 100 times. \textbf{Right:} Radial profiles of the weather and increased weather subtracted by no weather profile. While there is no real difference in results for the simulations with and without the weather phase screen, the results with the increased amplitude weather phase screen has a slight discrepancy, mostly on the wings of the PSF profile ($r>2.5$ pixels). 
    }
        \label{fig:Galsim_Radial}
\end{figure*}

Figure \ref{fig:Galsim_Field} shows the results from the simulation of 100 stars across the focal plane for the three different test cases. 
This shows that the `weather' does not affect any one region of the focal plane more than another but the overall PSF sigma is broader for the increased `weather' case. Averaging across all 100 stars yields an average PSF sigma of $1.0715 \pm 0.005$ for the `weather' scenario, $1.0710 \pm 0.005$ for the no `weather' scenario, and $1.129 \pm 0.005$ for the increased weather scenario. These would correspond to a seeing of $\sim 0.5$ arcseconds, which matches the best seeing conditions for Rubin Observatory \citep{Ivezic2019,LSSTRequierments}, consistent with our use of one airmass in the simulations.

For the case of a single star measured with 100 different instances of atmosphere and phase screen, we measure a PSF sigma of $1.072 \pm 0.001$ for the weather scenario, a PSF sigma of $1.072 \pm 0.001$ for the no-weather scenario, and a PSF sigma of $1.125 \pm 0.003$ for the increased weather scenario. We also measured the average radial profiles of the stars (Figure \ref{fig:Galsim_Radial}). The differences in the right panel of the figure, show that the wings of the PSF profile ($r>2.5$ are the most susceptible to change with change in the weather, as small changes in the pupil plane can impact the PSF shape at large distances, as can be seen in the right panel of Figure \ref{fig:Galsim_Radial}.

\section{Conclusion} 

In this paper, we discuss a turbulent `weather' pattern seen in EO testing of the LSSTCam. This was discovered evaluating difference images between flat field images and tracking their changes over time. We found evidence that this effect was affecting both flat differenced images and the 2-D correlation functions of the individual images. Further investigation found these changes were tied to the difference in index of refraction within the camera body and is affected by the VPC fan, which supplies air between the L3 and L2 lenses (air side). The sensitivity of our EO test data to this effect was due to the unique projection setup and short flash times of the illumination system that facilitated `freezing' the effect of density variations in the turbulent air and tracking the evolution of the effect.  
Using \textit{batoid} and \textit{galsim}, the PSF sigma was shown to only deteriorate by $10^{-4}$ due to this effect once the entire optical system was included via simulation.  This shows that the weather effect is minimized due to the fast optical system (f/1.2) and should not affect the LSST.
We were not able to find any literature about Camera purge system's effects on the PSF. And while this won't affect LSST, it might be useful to investigate this as telescopes and cameras continue to improve.

\section*{Disclosures}
The authors declare there are no financial interests, commercial affiliations, or other potential conflicts of interest that have influenced the objectivity of this research or the writing of this paper

\section*{Data Availability}
As the camera is currently being commissioned, image data is restricted to LSST/Rubin Users. 
\section*{Acknowledgments}
\begin{acknowledgments}
We would like to thank Tony Tyson, Chris Waters, Seth Digel Pierre Astier, and Pierre Antilogus, and the anonymous reviewers for insightful discussion and comments. This material is based upon work supported in part by the National Science Foundation through Cooperative
Agreement 1258333 managed by the Association of Universities for Research in Astronomy (AURA), and the
Department of Energy under Contract No. DE-AC02-76SF00515 with the SLAC National Accelerator Laboratory. Additional Rubin funding comes from private donations, grants to universities, and in-kind support from LSST-DA Institutional Members.
\end{acknowledgments}
\bibliography{main}{}

\newcommand{\noop}[1]{}
\begin{thebibliography}{}
\expandafter\ifx\csname natexlab\endcsname\relax\def\natexlab#1{#1}\fi
\providecommand{\url}[1]{\href{#1}{#1}}
\providecommand{\dodoi}[1]{doi:~\href{http://doi.org/#1}{\nolinkurl{#1}}}
\providecommand{\doeprint}[1]{\href{http://ascl.net/#1}{\nolinkurl{http://ascl.net/#1}}}
\providecommand{\doarXiv}[1]{\href{https://arxiv.org/abs/#1}{\nolinkurl{https://arxiv.org/abs/#1}}}

\bibitem[{{Astier} {et~al.}(2019){Astier}, {Antilogus}, {Juramy}, {Le Breton}, {Le Guillou}, \& {Sepulveda}}]{Astier2019}
{Astier}, P., {Antilogus}, P., {Juramy}, C., {et~al.} 2019, \aap, 629, A36, \dodoi{10.1051/0004-6361/201935508}

\bibitem[{{Bosch} {et~al.}(2019){Bosch}, {AlSayyad}, {Armstrong}, {Bellm}, {Chiang}, {Eggl}, {Findeisen}, {Fisher-Levine}, {Guy}, {Guyonnet}, {Ivezi{\'c}}, {Jenness}, {Kov{\'a}cs}, {Krughoff}, {Lupton}, {Lust}, {MacArthur}, {Meyers}, {Moolekamp}, {Morrison}, {Morton}, {O'Mullane}, {Parejko}, {Plazas}, {Price}, {Rawls}, {Reed}, {Schellart}, {Slater}, {Sullivan}, {Swinbank}, {Taranu}, {Waters}, \& {Wood-Vasey}}]{LSSTPipelines}
{Bosch}, J., {AlSayyad}, Y., {Armstrong}, R., {et~al.} 2019, in Astronomical Society of the Pacific Conference Series, Vol. 523, Astronomical Data Analysis Software and Systems XXVII, ed. P.~J. {Teuben}, M.~W. {Pound}, B.~A. {Thomas}, \& E.~M. {Warner}, 521, \dodoi{10.48550/arXiv.1812.03248}

\bibitem[{{Broughton} {et~al.}(2024){Broughton}, {Utsumi}, {Plazas Malag{\'o}n}, {Waters}, {Lage}, {Snyder}, {Rasmussen}, {Marshall}, {Chiang}, {Murgia}, \& {Roodman}}]{Broughton2024}
{Broughton}, A., {Utsumi}, Y., {Plazas Malag{\'o}n}, A.~A., {et~al.} 2024, \pasp, 136, 045003, \dodoi{10.1088/1538-3873/ad3aa2}

\bibitem[{{Esteves} {et~al.}(2023){Esteves}, {Utsumi}, {Snyder}, {Schutt}, {Broughton}, {Trbalic}, {Mau}, {Rasmussen}, {Plazas Malag{\'o}n}, {Bradshaw}, {Marshall}, {Digel}, {Chiang}, {Rykoff}, {Waters}, {Soares-Santos}, \& {Roodman}}]{Esteves2023}
{Esteves}, J.~H., {Utsumi}, Y., {Snyder}, A., {et~al.} 2023, \pasp, 135, 115003, \dodoi{10.1088/1538-3873/ad0a73}

\bibitem[{{Hirata} \& {Seljak}(2003)}]{2003MNRAS.343..459H}
{Hirata}, C., \& {Seljak}, U. 2003, \mnras, 343, 459, \dodoi{10.1046/j.1365-8711.2003.06683.x}

\bibitem[{{Ivezi{\'c}} \& the {LSST Science Collaboration}(2018)}]{LSSTRequierments}
{Ivezi{\'c}}, {\v{Z}}., \& the {LSST Science Collaboration}. 2018, The LSST System Science Requirements Document, Tech. rep., LSST-DA.
\newblock \url{https://docushare.lsst.org/docushare/dsweb/Get/LPM-17}

\bibitem[{{Ivezi{\'c}} {et~al.}(2019){Ivezi{\'c}}, {Kahn}, {Tyson}, {Abel}, {Acosta}, {Allsman}, {Alonso}, {AlSayyad}, {Anderson}, {Andrew}, {Angel}, {Angeli}, {Ansari}, {Antilogus}, {Araujo}, {Armstrong}, {Arndt}, {Astier}, {Aubourg}, {Auza}, {Axelrod}, {Bard}, {Barr}, {Barrau}, {Bartlett}, {Bauer}, {Bauman}, {Baumont}, {Bechtol}, {Bechtol}, {Becker}, {Becla}, {Beldica}, {Bellavia}, {Bianco}, {Biswas}, {Blanc}, {Blazek}, {Blandford}, {Bloom}, {Bogart}, {Bond}, {Booth}, {Borgland}, {Borne}, {Bosch}, {Boutigny}, {Brackett}, {Bradshaw}, {Brandt}, {Brown}, {Bullock}, {Burchat}, {Burke}, {Cagnoli}, {Calabrese}, {Callahan}, {Callen}, {Carlin}, {Carlson}, {Chandrasekharan}, {Charles-Emerson}, {Chesley}, {Cheu}, {Chiang}, {Chiang}, {Chirino}, {Chow}, {Ciardi}, {Claver}, {Cohen-Tanugi}, {Cockrum}, {Coles}, {Connolly}, {Cook}, {Cooray}, {Covey}, {Cribbs}, {Cui}, {Cutri}, {Daly}, {Daniel}, {Daruich}, {Daubard}, {Daues}, {Dawson}, {Delgado}, {Dellapenna}, {de Peyster}, {de Val-Borro}, {Digel}, {Doherty}, {Dubois},
  {Dubois-Felsmann}, {Durech}, {Economou}, {Eifler}, {Eracleous}, {Emmons}, {Fausti Neto}, {Ferguson}, {Figueroa}, {Fisher-Levine}, {Focke}, {Foss}, {Frank}, {Freemon}, {Gangler}, {Gawiser}, {Geary}, {Gee}, {Geha}, {Gessner}, {Gibson}, {Gilmore}, {Glanzman}, {Glick}, {Goldina}, {Goldstein}, {Goodenow}, {Graham}, {Gressler}, {Gris}, {Guy}, {Guyonnet}, {Haller}, {Harris}, {Hascall}, {Haupt}, {Hernandez}, {Herrmann}, {Hileman}, {Hoblitt}, {Hodgson}, {Hogan}, {Howard}, {Huang}, {Huffer}, {Ingraham}, {Innes}, {Jacoby}, {Jain}, {Jammes}, {Jee}, {Jenness}, {Jernigan}, {Jevremovi{\'c}}, {Johns}, {Johnson}, {Johnson}, {Jones}, {Juramy-Gilles}, {Juri{\'c}}, {Kalirai}, {Kallivayalil}, {Kalmbach}, {Kantor}, {Karst}, {Kasliwal}, {Kelly}, {Kessler}, {Kinnison}, {Kirkby}, {Knox}, {Kotov}, {Krabbendam}, {Krughoff}, {Kub{\'a}nek}, {Kuczewski}, {Kulkarni}, {Ku}, {Kurita}, {Lage}, {Lambert}, {Lange}, {Langton}, {Le Guillou}, {Levine}, {Liang}, {Lim}, {Lintott}, {Long}, {Lopez}, {Lotz}, {Lupton}, {Lust}, {MacArthur}, {Mahabal},
  {Mandelbaum}, {Markiewicz}, {Marsh}, {Marshall}, {Marshall}, {May}, {McKercher}, {McQueen}, {Meyers}, {Migliore}, {Miller}, {Mills}, {Miraval}, {Moeyens}, {Moolekamp}, {Monet}, {Moniez}, {Monkewitz}, {Montgomery}, {Morrison}, {Mueller}, {Muller}, {Mu{\~n}oz Arancibia}, {Neill}, {Newbry}, {Nief}, {Nomerotski}, {Nordby}, {O'Connor}, {Oliver}, {Olivier}, {Olsen}, {O'Mullane}, {Ortiz}, {Osier}, {Owen}, {Pain}, {Palecek}, {Parejko}, {Parsons}, {Pease}, {Peterson}, {Peterson}, {Petravick}, {Libby Petrick}, {Petry}, {Pierfederici}, {Pietrowicz}, {Pike}, {Pinto}, {Plante}, {Plate}, {Plutchak}, {Price}, {Prouza}, {Radeka}, {Rajagopal}, {Rasmussen}, {Regnault}, {Reil}, {Reiss}, {Reuter}, {Ridgway}, {Riot}, {Ritz}, {Robinson}, {Roby}, {Roodman}, {Rosing}, {Roucelle}, {Rumore}, {Russo}, {Saha}, {Sassolas}, {Schalk}, {Schellart}, {Schindler}, {Schmidt}, {Schneider}, {Schneider}, {Schoening}, {Schumacher}, {Schwamb}, {Sebag}, {Selvy}, {Sembroski}, {Seppala}, {Serio}, {Serrano}, {Shaw}, {Shipsey}, {Sick}, {Silvestri},
  {Slater}, {Smith}, {Smith}, {Sobhani}, {Soldahl}, {Storrie-Lombardi}, {Stover}, {Strauss}, {Street}, {Stubbs}, {Sullivan}, {Sweeney}, {Swinbank}, {Szalay}, {Takacs}, {Tether}, {Thaler}, {Thayer}, {Thomas}, {Thornton}, {Thukral}, {Tice}, {Trilling}, {Turri}, {Van Berg}, {Vanden Berk}, {Vetter}, {Virieux}, {Vucina}, {Wahl}, {Walkowicz}, {Walsh}, {Walter}, {Wang}, {Wang}, {Warner}, {Wiecha}, {Willman}, {Winters}, {Wittman}, {Wolff}, {Wood-Vasey}, {Wu}, {Xin}, {Yoachim}, \& {Zhan}}]{Ivezic2019}
{Ivezi{\'c}}, {\v{Z}}., {Kahn}, S.~M., {Tyson}, J.~A., {et~al.} 2019, \apj, 873, 111, \dodoi{10.3847/1538-4357/ab042c}

\bibitem[{{Janesick}(2001)}]{Janesick2001}
{Janesick}, J.~R. 2001, {Scientific charge-coupled devices}

\bibitem[{{Jarvis} {et~al.}(2004){Jarvis}, {Bernstein}, \& {Jain}}]{Jarvis2004}
{Jarvis}, M., {Bernstein}, G., \& {Jain}, B. 2004, \mnras, 352, 338, \dodoi{10.1111/j.1365-2966.2004.07926.x}

\bibitem[{{Kotov} {et~al.}(2016){Kotov}, {Haupt}, {O'Connor}, {Smith}, {Takacs}, {Neal}, \& {Chiang}}]{Kotov2016}
{Kotov}, I.~V., {Haupt}, J., {O'Connor}, P., {et~al.} 2016, in Society of Photo-Optical Instrumentation Engineers (SPIE) Conference Series, Vol. 9915, High Energy, Optical, and Infrared Detectors for Astronomy VII, ed. A.~D. {Holland} \& J.~{Beletic}, 99150V, \dodoi{10.1117/12.2231925}

\bibitem[{{Lopez} {et~al.}(2018){Lopez}, {Marshall}, {Bond}, {Haupt}, {Johnson}, {Neal}, {O'Connor}, {Rasmussen}, {Roodman}, {Takacs}, \& {Utsumi}}]{Lopez2018}
{Lopez}, M., {Marshall}, S., {Bond}, T., {et~al.} 2018, in Society of Photo-Optical Instrumentation Engineers (SPIE) Conference Series, Vol. 10702, Ground-based and Airborne Instrumentation for Astronomy VII, ed. C.~J. {Evans}, L.~{Simard}, \& H.~{Takami}, 107022C, \dodoi{10.1117/12.2312200}

\bibitem[{{Mandelbaum} {et~al.}(2005){Mandelbaum}, {Hirata}, {Seljak}, {Guzik}, {Padmanabhan}, {Blake}, {Blanton}, {Lupton}, \& {Brinkmann}}]{Mandelbaum2005}
{Mandelbaum}, R., {Hirata}, C.~M., {Seljak}, U., {et~al.} 2005, \mnras, 361, 1287, \dodoi{10.1111/j.1365-2966.2005.09282.x}

\bibitem[{Meyers {et~al.}(2019)Meyers, Kirkby, \& Thomas}]{Batoid}
Meyers, J.~E., Kirkby, D., \& Thomas, D. 2019, batoid, [Computer Software] \url{https://doi.org/10.11578/dc.20200708.1}, \dodoi{10.11578/dc.20200708.1}

\bibitem[{{Newbry} {et~al.}(2018){Newbry}, {Lange}, {Roodman}, {Reil}, {Bond}, {Rasmussen}, {Bowdish}, {Snyder}, {Rosenberg}, \& {Lee}}]{Newbry2018}
{Newbry}, S., {Lange}, T., {Roodman}, A., {et~al.} 2018, in Society of Photo-Optical Instrumentation Engineers (SPIE) Conference Series, Vol. 10702, Ground-based and Airborne Instrumentation for Astronomy VII, ed. C.~J. {Evans}, L.~{Simard}, \& H.~{Takami}, 1070258, \dodoi{10.1117/12.2314269}

\bibitem[{{O'Connor} {et~al.}(2016){O'Connor}, {Antilogus}, {Doherty}, {Haupt}, {Herrmann}, {Huffer}, {Juramy-Giles}, {Kuczewski}, {Russo}, {Stubbs}, \& {Van Berg}}]{OConnor2016}
{O'Connor}, P., {Antilogus}, P., {Doherty}, P., {et~al.} 2016, in Society of Photo-Optical Instrumentation Engineers (SPIE) Conference Series, Vol. 9915, High Energy, Optical, and Infrared Detectors for Astronomy VII, ed. A.~D. {Holland} \& J.~{Beletic}, 99150X, \dodoi{10.1117/12.2232729}

\bibitem[{{Plazas Malag{\'o}n} {et~al.}(2024){Plazas Malag{\'o}n}, {Waters}, {Broughton}, {Rykoff}, {Fert{\'e}}, {Fisher-Levine}, \& {Lupton}}]{Plazas2024}
{Plazas Malag{\'o}n}, A.~A., {Waters}, C., {Broughton}, A., {et~al.} 2024, arXiv e-prints, arXiv:2404.14516, \dodoi{10.48550/arXiv.2404.14516}

\bibitem[{{Roodman} {et~al.}(\noop{3001}in press){Roodman}, {Rasmussena}, {Bradshaw}, {Charles}, {Digel}, {Duboisa}, {Johnson}, {Kahn}, {Liang}, {Marshall}, {Neal}, {Reil}, {Rykoff}, {Schindler}, {Schutt}, {Utsumi}, {Bogart}, {Bond}, {Bowdish}, {Cisneros}, {Eisner}, {Freytag}, {Hascall}, {Lange}, {Lazarte}, {Lopez}, {Mendez}, {Plazas Malagon}, {Newbry}, {Nordby}, {Onoprienkoa}, {Osier}, {Pollek}, {Qiu}, {Saxton}, {Tether}, {Thayer}, {Turri}, {Banovetz}, {O’Connor}, {Riot}, {Wolfe}, {Lage}, {Polin}, {Snyder}, {Tyson}, {Nichols}, {Ritz}, {Shestakov}, {Wood}, {Broughton}, {Park}, {Esteves}, {Barrau}, {Bregeon}, {Combet}, {Dargaud}, {Lagorio}, {Migliore}, {Vezzu}, {Antilogus}, {Astier}, {Daubard}, {Juramy}, {Laporte}, {Guillemin,}, {Aubourg}, {Boucaud}, {Parisel}, {Virieux}, {Breugnon}, {Karst}, {Marini}, {Fisher-Levine}, \& {Waters}}]{Roodman2024}
{Roodman}, A., {Rasmussena}, A., {Bradshaw}, A., {et~al.} \noop{3001}in press, in Society of Photo-Optical Instrumentation Engineers (SPIE) Conference Series, Vol. 13096, Ground-based and Airborne Instrumentation for Astronomy X

\bibitem[{{Rowe} {et~al.}(2015){Rowe}, {Jarvis}, {Mandelbaum}, {Bernstein}, {Bosch}, {Simet}, {Meyers}, {Kacprzak}, {Nakajima}, {Zuntz}, {Miyatake}, {Dietrich}, {Armstrong}, {Melchior}, \& {Gill}}]{Rowe2015}
{Rowe}, B.~T.~P., {Jarvis}, M., {Mandelbaum}, R., {et~al.} 2015, Astronomy and Computing, 10, 121, \dodoi{10.1016/j.ascom.2015.02.002}

\bibitem[{{Snyder} {et~al.}(2021){Snyder}, {Longley}, {Lage}, {Digel}, {Neal}, {Utsumi}, \& {Roodman}}]{Synder2021}
{Snyder}, A., {Longley}, E., {Lage}, C., {et~al.} 2021, Journal of Astronomical Telescopes, Instruments, and Systems, 7, 048002, \dodoi{10.1117/1.JATIS.7.4.048002}

\bibitem[{{Snyder} {et~al.}(2020){Snyder}, {Barrau}, {Bradshaw}, {Bowdish}, {Chiang}, {Combet}, {Digel}, {Dubois}, {Eraud}, {Juramy}, {Lage}, {Lange}, {Migliore}, {Nomerotski}, {O'Connor}, {Park}, {Rasmussen}, {Reil}, {Roodman}, {Shestakov}, {Utsumi}, \& {Wood}}]{Snyder2020}
{Snyder}, A., {Barrau}, A., {Bradshaw}, A., {et~al.} 2020, in Society of Photo-Optical Instrumentation Engineers (SPIE) Conference Series, Vol. 11454, X-Ray, Optical, and Infrared Detectors for Astronomy IX, ed. A.~D. {Holland} \& J.~{Beletic}, 1145439, \dodoi{10.1117/12.2562915}

\bibitem[{{Ustumi} {et~al.}(in press){Ustumi}, {Antilogus}, {Astier}, {Banovetz}, {Bradshaw}, {Bregeon}, {Broughton}, {Chiang}, {Combet}, {Dargaud}, {Digel}, {Esteves}, {Guillemin}, {Johnson}, {Juramy}, {Lage}, {Liang}, Stuart, {Migliore}, {Neal}, {Nichols}, {Polin}, {Rasmussen}, Steve, Eli, {Roodman}, {Schutt}, {Sheshtakov}, {Snyder}, {Thayer}, {Turri}, {Tyson}, \& Duncan}]{Utsumi2024}
{Ustumi}, Y., {Antilogus}, P., {Astier}, P., {et~al.} in press, in Society of Photo-Optical Instrumentation Engineers (SPIE) Conference Series, Vol. 13103, X-Ray, Optical, and Infrared Detectors for Astronomy XI

\end{thebibliography}


%






\appendix
Below are the parameters used to create the atmosphere and stellar sources for the input of \texttt{batoid}.
\begin{lstlisting}[language=Python]
atm = galsim.Atmosphere(
    1024.0,
    rng=rng,
    r0_500=0.2, 
    L0=25.0,
    r0_weights=[0.3, 0.3, 0.2, 0.1, 0.05, 0.05],
    direction=[ud()*galsim.degrees for _ in range(6)],
    speed=[ud()*20 for _ in range(6)]
)
atm_psf = atm.makePSF(626, diam=8.36)

rotTelPos = 35.0*galsim.degrees
rotSkyPos = 10.0*galsim.degrees
q = rotTelPos - rotSkyPos
lam=626
diam=8.36
scale_unit=galsim.arcsec
star=galsim.Airy(lam=lam,diam=diam,scale_unit=scale_unit)
sed = galsim.SED("vega.txt", wave_type='nm', flux_type='flambda')
bp = galsim.Bandpass("LSST_r.dat", wave_type='nm')
surface_ops = [
    galsim.WavelengthSampler(sed, bp, rng=rng),
    galsim.PhotonDCR(
        626, 
        zenith_angle=0*galsim.degrees,
        parallactic_angle=q
    )
]
star_psf = galsim.Convolve(
    star,
    atm.makePSF(
        626, 
        diam=8.36,
        obscuration=0.606,
    )
)
\end{lstlisting}

\end{document}